\documentclass[final, 5p, times, twocolumn]{elsarticle}

\usepackage[english]{babel}
\usepackage{amsmath}  % Math
\usepackage{booktabs}
\usepackage{graphicx} % Figures
\usepackage{adjustbox}% Resizing
\usepackage{breqn}    
\usepackage{minted}

\usepackage[colorlinks=true, allcolors=blue]{hyperref} 
\usepackage{cleveref}

\begin{document}

\begin{frontmatter}

\title{\textbf{Leveraging Cellular Automata for Real-Time Wildfire Spread Modeling in California}}

\author[cross]{Connor Weinhouse\corref{cor1}}
\cortext[cor1]{Corresponding author}
\ead{connorweinhouse@gmail.com}
\author[uga]{Jameson Augustin}

\address[cross]{High School Student at the Crossroads School for Arts \& Sciences}
\address[uga]{Department of Agricultural and Applied Economics, University of Georgia, Athens, GA 30602 USA}

\begin{abstract}
Wildfires are becoming increasingly frequent and devastating, and therefore the technology to combat them must adapt accordingly. Modern predictive models have failed to balance predictive accuracy and operational viability, resulting in consistently delayed or misinformed fire suppression and public safety efforts. The present study addresses this gap by developing and validating a predictive model based on cellular automata (CA) that incorporates key environmental variables, including vegetation density (NDVI), wind speed and direction, and topographic slope derived from open-access datasets. The presented CA framework offers a lightweight alternative to data-heavy approaches that fail in emergency contexts. Evaluation of the model using a confusion matrix against burn scars from the 2025 Pacific Palisades Fire yielded a recall of 0.860, a precision of 0.605, and an overall F1 score of 0.711 after 50 parameter optimization trials, with each simulation taking an average of 1.22 seconds. CA-based models can bridge the gap between accuracy and applicability, successfully guiding public safety and fire suppression efforts.
\end{abstract}


\end{frontmatter}

\section{Introduction}
As climate change has worsened in recent years, wildfires have become more common and disastrous, especially in dry, high-wind areas such as California \citep{CDFW_wildfire_impacts}. The past five years have witnessed the devastation of California by five of the largest wildfires in history, and globally, fires are becoming increasingly disastrous and frequent \citep{calfire2020, epa_wildfires}. The Palisades Fire in January of 2025  resulted in the destruction of 7,000 structures and death of 12 people, a permanent set back for the Palisades community \citep{calfire2025palisades}.

This research has three primary goals: (1) to design a cellular automata (CA) model that simulates the spread of wildfire and validate the simulated area against an observed burn scar;\footnote{Cellular Automata are computational models that define an area as a grid of cells, each with an inherent state that can change over time steps according to pre-determined rules.} (2) to incorporate environmental inputs from satellite data, including vegetation density (NDVI), wind speed and strength, and topographic slope; and (3) to optimize the model’s parameters, finally examining its ability to inform fire suppression efforts.

Past research in wildfire modeling consistently fails due to its incorrect priority of accuracy over operational usability. Two clear examples of this are the SeasFire Cube and the Firemap web platform \citep{crawl2017firemap, karasante2025seasfire}; both rely on dozens of environmental and human inputs, ultimately sacrificing their models' usability in emergency circumstances for high accuracy. More recent approaches have used CA frameworks, such as the modeling proposed by Xu et al., which derived cell transition rules from deep learning, but their use of machine learning makes them highly resource-intensive, as also seen in many other studies \citep{zou2023attention, xu2022modeling}.

Emergency responders' public safety efforts in California have recently been informed by the Fire Integrated Real-Time Intelligence System (FIRIS), which relies on near-real-time modeling from the WIFIRE Firemap, and other predictions from AI-driven models developed by Southern California Edison (SCE) \citep{UCSD_FIRIS, SCE2025WMP}. These provide emergency responders with predictions on the direction and rate of the fire's spread, allowing firefighters to use the simulation to inform their suppression efforts, but are highly computationally intensive, inaccurate, and slow to update in long-lasting wildfires \citep{Altintas2025Firemap}. This was painfully seen in the Eaton and Palisades Fires of January 2025, in which SCE underestimated the potential of the fires by a factor of ten, ultimately misguiding suppression efforts \citep{McLaughlin2025EdisonWildfireForecast}.

Collectively, these studies demonstrate a clear gap in the consistent sacrifice of the reactive speed of current models for unnecessary levels of precision. Our research seeks to bridge this gap with a novel model that balances physical accuracy and real-time applicability by evaluating a CA-based fire spread model grounded in real-world data and practical application. This structure is promising for applications that require rapid feedback under variable environmental conditions.

The paper is structured as follows: Section \ref{sec:litreview} reviews past research in the field and its oversights. Section \ref{sec:data} details the data and their sources used in the study. Section \ref{sec:methods} outlines the methods employed. Section \ref{sec:results} evaluates the results of the simulation against the observed burn area. Section \ref{sec:discussions} discusses the implications of our findings. Section \ref{sec:conclusions} concludes with limitations, contributions, and future research paths.

\section{Literature Review}\label{sec:litreview}
As the effects of climate change worsen, predictive wildfire spread technology is urgently needed to address the ever-increasing rate of fires worldwide. Although technology exists, models that cater to real-time emergency response represent a gap in the field and must be prioritized. Fire behavior has been modeled from countless angles since the early work of Rothermel and colleagues, but spatial resolution, computational efficiency, and environmental integration remain a challenge \citep{rothermel1972mathematical}. Notably seen in the more than 30,000 damaged or destroyed structures and the multibillion-dollar economic impact in the Los Angeles wildfires of January 2025, inadequate fire prediction results in drastic devastation to wildlife and urban communities \citep{laedc_wildfires_2025}. No single approach has resulted in a model that is simultaneously accurate, data-driven, and operationally viable in real-time emergency contexts. This literature review examines key contributions to the field to explain the intent behind the CA-based approach in the present study, highlighting progress in environmental data integration, reactive modeling, and satellite-based detection, while addressing the ongoing need for models that are both precise and operationally viable.

Environmental variables are the basis of wildfire models in their use to inform landscape and atmospheric conditions. Karasante et al. recently developed the SeasFire Cube,\footnote{The SeasFire Cube is a complex scientific datacube designed to support global wildfire prediction.} a primary example of a large-scale multivariate dataset that uses 59 environmental and anthropogenic variables to offer improved understanding and anticipation of wildfires, but its spatial and temporal resolution limits its real-time applicability \citep{karasante2025seasfire}. High-resolution hyperspectral data have also been used to produce fire temperature and burn severity maps after the 2003 Simi Fire, illustrating the integral role remote imaging plays in fire research and representing the focus of early literature on burn analysis rather than predictive modeling \citep{dennison2006wildfire}. Complementing the two, Or and colleagues examined the role of soil properties in fire ignition and propagation \citep{or2023review}. These studies show the evident progression of fire information technology toward highly integrated environmental systems. Models dependent on large-scale multivariate datasets, such as SeasFire Cube, sacrifice real-time response for analytical global coverage, while models focusing on rapid response time sacrifice accuracy and variable input.

Modern modeling approaches have been characterized by a combination of CA and machine learning frameworks designed to improve simulation accuracy while limiting computational complexity. The study by Xu et al. utilizes CA and adopts fire state transition rules from real-world datasets through machine learning, teaching the system to learn from environmental variables rather than hard-coding optimized parameters \citep{xu2022modeling}. A more deep learning-oriented approach was taken by Zou et al. using a convolutional neural network (CNN) model trained on fire tracking satellites and basic environmental inputs such as wind, fuel, and temperature \citep{zou2023attention}. Their approach relies too heavily on high-resolution remote imaging and complex model training, limiting its viability for immediate emergency response. The Firemap web platform is a highly deliberate and calculated model,\footnote{Firemap is a part of the Workflows Integrating Collaborative Hazard Sciences (WIFIRE) cyberinfrastructure project. It uses emerging AI techniques and data input to produce public dynamic visualizations of fire spread.} accounting for numerous diverse inputs such as remote imaging, sensor networks, and even social media \citep{crawl2017firemap}. Firemap's philosophy focuses on responsiveness and real-world data, but its shortcomings are evident in its complexity and data-heavy nature. The intricacy and infrastructural demands of the system encounter operational barriers when necessary in emergency situations.

These studies illustrate a shared convergence on the use of complex multivariate environmental data, remote imaging, and high computational demand in wildfire modeling. However, a key gap is clear in the need for rapid, easily interpretable, and validated fire spread model that can operate in low-computation environments and be deployed in active fire events. The field is shifting to machine learning approaches, yet this only increases complexity with limited returns on predictive accuracy. Furthermore, few studies offer validation against recent impactful fire events. Addressing this critical gap, the present study develops and evaluates a CA model that integrates the key variables of NDVI, wind speed and direction, and elevation to model fire spread with validation against recent burn scars from the January 2025 Palisades Fires.

\section{Data}\label{sec:data}
\subsection{Data Sources}
This study integrates multivariate geospatial datasets to model and validate the spread of the wildfire during the recent January 2025 Palisades Fire. The study relies on four primary datasets: (1) the Copernicus Sentinel-2 mission supplies vegetation density for burn severity mapping and NDVI; (2) the Visible Infrared Imaging Radiometer Suite (VIIRS) to identify the ignition point of the fire and verify its progression; (3) the ERA5 atmospheric reanalysis dataset offers wind direction and speed;\footnote{ERA5 is the fifth generation of atmospheric reanalysis by the European Centre for Medium-Range Weather Forecasts (ECMWF)} and (4) the Copernicus Digital Elevation Model (DEM) to derive the topographic slope of the area of interest. These variables were later used to define a realistic burn probability for simulation and to validate the produced scar against the actual burn scar. All datasets used in this study are publicly available, ensuring that the described methodology can be extended by other researchers.

\subsection{Study Area}
The dedicated area of interest for the current study is the Pacific Palisades in Southern California, a tightly knit coastal community with a long history of fires. The Palisades' dry climate and abundant vegetation of the Santa Monica mountains, the strength and volatility of the Santa Ana winds, and the encroaching threat of climate change make the area extremely susceptible to unexpected wildfires.

\begin{figure}[H]
    \centering
    \includegraphics[width=1.0\columnwidth]{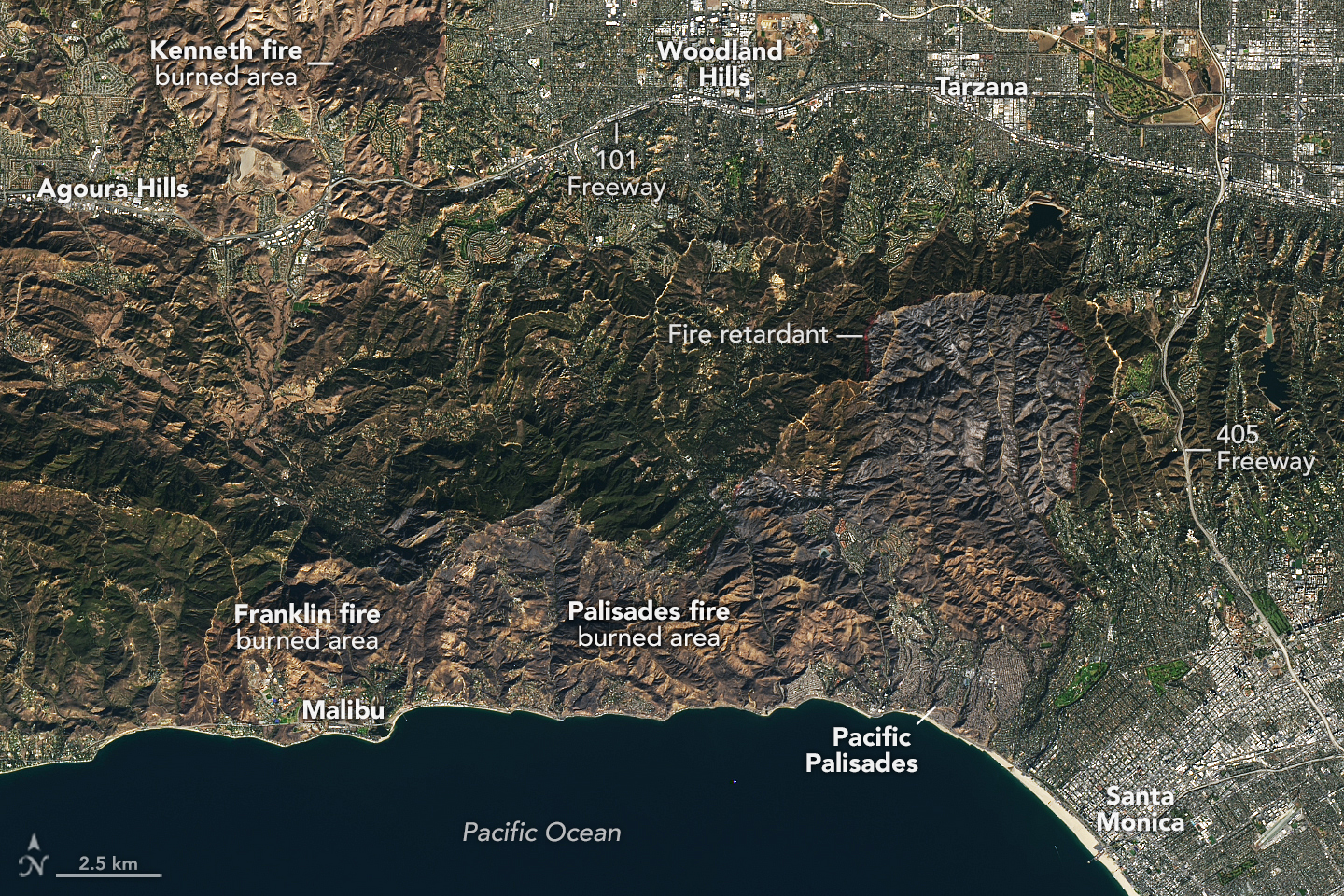} 
    \caption{The Palisades Fire Footprint}
    \label{fig:earth}
\end{figure}

The Palisades endured an extremely destructive and painful wildfire in January of 2025 as presented in \autoref{fig:earth}.\footnote{This map was retrieved from the NASA Earth Observatory website: \url{https://earthobservatory.nasa.gov/images/153831/the-palisades-fires-footprint}} Over the course of just 24 days, the fire burned 23,448 acres, destroyed 6,837 structures, and killed 12 people \citep{calfire2025palisades}. The fire's poor evacuation orders in conjunction with the irregularity of the climate and speed of the Santa Ana winds had a devastating community impact.

% add a map of the area here

\subsection{Satellite Imagery}
\subsubsection{Sentinel-2}
The Copernicus Sentinel-2 mission from the European Space Agency was the main dataset used in the present study.\footnote{Copernicus Sentinel-2 Data Source: \url{https://dataspace.copernicus.eu/explore-data/data-collections/sentinel-data/sentinel-2}} Both the analysis and simulation used Level-2A (L2A) surface reflectance data to offer atmospherically corrected imaging. Sentinel-2's satellites provide high-resolution optical images and 13 spectral bands to facilitate pre- and post-fire landscape condition analysis. The study used the following spectral bands for this analysis:
\begin{itemize}
    \item Blue (B2)
    \item Green (B3)
    \item Red (B4)
    \item Near-Infrared (NIR, B8)
    \item Shortwave-Infrared (SWIR, B12)
\end{itemize}
Additionally, Sentinel-2's Scene Classification Layer (SCL) allowed for atmospheric correction, and masking of cloud and water bodies prior to further usage of data, ensuring all analysis was centered on vegetation shifts. 
\subsubsection{VIIRS}
The study also used remote imaging data from the VIIRS to identify the ignition point and track the fire's spread.\footnote{VIIRS Active Fire Data Source: \url{https://earthdata.nasa.gov/earth-observation-data/near-real-time/firms/viirs-i-band-active-fire-data}} This NASA sensor collects visible and infrared imagery as well as global observations of Earth from onboard the Suomi NPP and JPSS satellites. Of its many functions, the VIIRS provides active fire detections that are fundamental for the initialization of the fire spread simulation and for validating the fire's early progression.
\subsection{Meteorological Data}
The study incorporated meteorological data from the ERA5 atmospheric reanalysis dataset produced by the European Centre for Medium-Range Weather Forecasts (ECMWF) and accessed through the Copernicus Climate Data Store.\footnote{ERA5 Meteorological Data Source: \url{https://cds.climate.copernicus.eu/datasets/reanalysis-era5-single-levels?tab=download}} ERA5 provides hourly estimates of a multitude of atmospheric variables at high spatial resolution. For the purposes of our study, we used the 10-meter u-component (east-west) and v-component (north-south) of the wind to calculate wind speed and direction to influence the directional behavior in our simulation. Wind speed, arguably the most crucial variable in fires, and wind direction are essential to predict the rate and direction of fire spread.
\subsection{Topographic Data}
Lastly, the study used topographic data from the Copernicus DEM. Due to the diverse elevations present throughout Pacific Palisades and the surrounding area, elevation played a key role in the fire's spread. We used the public GLO-30 product, which offers global coverage at a resolution of 30 meters.\footnote{Copernicus DEM GLO-30 Data Source: \url{https://dataspace.copernicus.eu/explore-data/data-collections/copernicus-contributing-missions/collections-description/COP-DEM}} With this data, we were able to calculate the topographic slope of the burn region, a significant parameter in the fire spread model. 

\section{Methods}\label{sec:methods}

\subsection{Burnt Area Mapping}
The entirety of the 23,448 acres burned over the course of 24 days in the Pacific Palisades fire was determined using the dNBR, a remote sensing spectral index widely used to analyze burn severity \citep{calfire2025palisades}. Our area of interest for the simulation was designated to the bounding box coordinates: [-118.73999, 34.01176, -118.48914, 34.14894].

\subsubsection{Normalized Burn Ratio (NBR)}
The NBR is calculated using the Near-Infrared (NIR) and Shortwave-Infrared (SWIR) bands derived from the Sentinel-2 data. It is calculated using this formula:
\begin{equation}
    \text{NBR} = \frac{\text{NIR} - \text{SWIR}}{\text{NIR} + \text{SWIR}}
\end{equation}
Vegetation consistently reflects values of NIR and SWIR correlating with its health. Healthy vegetation reflects high NIR and low SWIR reflectance, resulting in a high NBR value, while burned areas exhibit low NIR reflectance and high SWIR reflectance, leading to a low NBR value.
\subsubsection{Differenced Normalized Burn Ratio (dNBR)}
To quantify and analyze the impact of the fire, we calculated the dNBR by subtracting the post-fire NBR from the pre-fire NBR: 
\begin{equation}
    \text{dNBR} = \text{NBR}_{\text{pre-fire}} - \text{NBR}_{\text{post-fire}}
\end{equation}
A larger dNBR value indicates a greater impact of the fire. Prior to the major fire in January 2025, a large section of our area of interest had been burned in a previous fire. This resulted in a very low NBR before the fire in January 2025, demonstrating the importance of dNBR for representing the impact rather than simply exhibiting its effects.
\subsubsection{Preprocessing}
Before analyzing the impact of the fire, all pre- and post-fire data were masked to remove clouds and water bodies using the Sentinel-2 SCL, which identifies pixels corresponding to shadows, clouds, and bodies of water. These masked regions were excluded from the data to ensure accurate spectral indices. After calculating dNBR, the resulting map was thresholded using a value of 0.3 to categorize pixels as ``burned" or ``unburned". This threshold was selected based on its common use in fire severity analysis studies and visual inspection of the resulting fire mask when the threshold was changed. The resulting dNBR map was then thresholded to create a binary burn mask. Finally, to refine the burn mask, a series of morphological operations were applied. Binary closing was used to fill small, inaccurate holes in the burn area, followed by erosion and dilation to remove insignificant isolated patches.
\subsection{Fire Spread Simulation}
The spread of the wildfire was simulated through a CA model. This approach models the area of interest as a grid of cells, with each cell representing a 10-meter by 10-meter square aligning with the resolution of the Sentinel-2 data. Each cell is assigned a state that evolves based on the states of its neighbors and predefined rules and probabilities. We chose to use a CA model for its balance between computational efficacy and spatial realism. Unlike complex neural networks and deep learning models as used in the attention-based modeling proposed by Zou et al. and the intricate multivariate SeasFire cube designed by Karasante et al., the CA model offers an easily interpretable, simple alternative without sacrificing model accuracy \citep{karasante2025seasfire, zou2023attention}.
\subsubsection{Cellular Automaton States}
The foundation of a CA model is its varying cell states, with states being able to transition based on defined rules and probabilities. Each cell in the simulation grid can be in one of these four states:
\begin{itemize}
    \item \textbf{Flammable:} The cell contains vegetation that can burn.
    \item \textbf{Burning:} The cell is presently on fire.
    \item \textbf{Burnt:} The cell has burned and can no longer burn.
    \item \textbf{Non-flammable:} The cell represents a water body or bare ground and cannot burn.
\end{itemize}
\subsubsection{State Transition Rules}
The simulation has a predefined number of time steps, with each time step representing a set interval of time while the simulation is running. These time steps allow the model to simulate the progression of the fire in a logical manner as well as the result. Each cell's state is updated at every time step based on the following predefined rules:
\begin{enumerate}
    \item A burning cell becomes a burnt cell after one time step.
    \item A burnt cell remains burnt.
    \item A nonflammable cell remains nonflammable.
    \item A flammable cell can ignite if it is adjacent to a burning cell, with a probability determined by the burn probability, $p_{\text{burn}}$.
\end{enumerate}

\subsubsection{Burn Probability}
The probability of the ignition of a flammable cell is a function of vegetation, topographic slope, and wind derived from previously mentioned remote imaging. The burn probability is based off of this formula:
\begin{equation}
    p_{\text{burn}} = p_0 \cdot f_{\text{NDVI}} \cdot f_{\text{slope}} \cdot f_{\text{wind}}
\end{equation}
where:
\begin{itemize}
    \item $p_0$ is the base ignition probability.
    \item $f_{\text{NDVI}}$ is a factor modifying the probability based on vegetation density.
    \item $f_{\text{slope}}$ is a factor accounting for the effect of topography.
    \item $f_{\text{wind}}$ is a factor representing the influence of wind.
\end{itemize}
Each factor is scaled to a value between 0 and 1 to ensure $p_{\text{burn}}$ remains within a valid range.
\paragraph{Vegetation Modifier ($f_{\text{NDVI}}$):}
The vegetation modifier is determined by the NDVI of the cell. The NDVI assigns a value to a cell based on vegetation health and density, with -1 quantifying no vegetation and +1 corresponding to highly flammable vegetation. Cells are classified into different thresholds, each with a specific modifier value that signifies their likelihood of ignition as shown below:
\begin{table}[ht]
    \centering
    \caption{NDVI-based Fuel Classification and $\mathbf{f_{\text{NDVI}}}$ Multiplicative Penalty Values. $T_W=0.1$ (Water Threshold), $T_L=0.4$ (Low Threshold), and $T_H=0.75$ (High Threshold).}
    \label{tab:ndvi_modifier}
    \begin{tabular}{llc}
    \toprule
    \textbf{Fuel Type} & \textbf{NDVI Range} & $\mathbf{f_{\text{NDVI}}}$ \textbf{Value} \\
    \midrule
    Water/Non-Flammable & $\text{NDVI} < 0.1$ & $-1$ \\
    Sparse Vegetation & $0.1 \leq \text{NDVI} < 0.4$ & $-1$ \\
    Medium Vegetation & $0.4 \leq \text{NDVI} < 0.75$ & $0.3$ \\
    Dense Vegetation & $\text{NDVI} \geq 0.75$ & $0.1$ \\
    \bottomrule
    \end{tabular}
\end{table}
\paragraph{Slope Modifier ($f_{\text{slope}}$):}
The large variability of topographic slope in the Pacific Palisades makes it a key modifier in understanding fire spread in our area of interest, as fires spread more rapidly uphill because the flames lay down into the slope, essentially pre-heating uphill vegetation \citep{sa_fire_behaviour_2025}. This effect on fire spread is modeled as an exponential function of the elevation difference between the burning and flammable cells:
\begin{equation}
    f_{\text{slope}} = e^{c_{\text{slope}} \cdot (\text{elevation}_{\text{flammable}} - \text{elevation}_{\text{burning}})}
\end{equation}
where $c_{\text{slope}}$ is a parameter that controls the strength of the slope effect. This formulation captures the tendency of fire to spread more rapidly uphill.

\paragraph{Wind Modifier ($f_{\text{wind}}$):}
Wind was another critical variable in the analysis and simulation of the Pacific Palisades fire due to its irregular role in the event. The dry Santa Ana winds reached abnormal speeds of 80 miles per hour early in the spread of the fire, a key reason behind its devastating impact \citep{cnn_wildfires_2025}. The wind modifier is also modeled as an exponential function, considering the component of the wind in the direction of fire spread:
\begin{equation}
    f_{\text{wind}} = e^{c_{\text{wind}} \cdot \text{wind\_component}}
\end{equation}
where $c_{\text{wind}}$ is a parameter controlling the wind's influence. A positive wind component (wind blowing from the burning cell to the flammable cell) increases the burn probability, while a negative component decreases it.
\subsection{Model Evaluation and Optimization}
The performance of the simulation was evaluated by validating the simulated burn scar against the observed burn area derived from the earlier dNBR analysis. The primary metric used for this evaluation was the F1 score.
\subsubsection{Confusion Matrix}
The success of the comparison between the simulated and observed burn areas is assessed in a confusion matrix, a table layout that allows for visualization of the performance of an algorithm, which includes the following components:
\begin{itemize}
    \item \textbf{True Positives (TP):} Pixels correctly identified as burned by the simulation.
    \item \textbf{False Positives (FP):} Pixels incorrectly identified as burned by the simulation.
    \item \textbf{False Negatives (FN):} Pixels that were burned in reality but were missed by the simulation.
    \item \textbf{True Negatives (TN):} Pixels correctly identified as unburned by the simulation.
\end{itemize}
\subsubsection{F1 Score}
The F1 score is the harmonic mean of precision and recall, providing a single metric to assess the simulation's overarching accuracy. It is calculated as follows:
\begin{equation}
    F1 = \frac{2 \cdot \text{TP}}{2 \cdot \text{TP} + \text{FP} + \text{FN}}
\end{equation}
F1 score is commonly used in wildfire modeling to evaluate simulations, and its balance between precision and recall makes it a useful quantification for success. An F1 score of 1 indicates a perfect match between the simulation and the observed data, while 0 represents a complete mismatch between the two.
\subsubsection{Parameter Optimization}
To find the optimal set of parameters for the simulation ($p_0$, $c_{\text{slope}}$, $c_{\text{wind}}$, and $f_{\text{NDVI}}$), a parameter optimization routine was employed. The optimization randomly sampled parameter values within predefined ranges based on their known role in fire spread as informed by prior studies. The simulation was run for each sampled set over 200 time steps, and the resulting scar was assessed using the F1 score. This process was repeated for 50 trials and the set of parameters that produced the highest F1 score was selected as the most optimal. These optimized parameters heightened spatial agreement between the simulated and observed burn areas, illustrating that fine-tuning the parameters is critical to accurately capture fire dynamics. This optimization approach offers an efficient way to calibrate the model to successfully display the influence of complex environmental factors on fire spread.
\section{Results}\label{sec:results}
\subsection{Observed Burn Scar}
\begin{figure}[H]
    \centering
    
        \centering
        \includegraphics[width=1.0\columnwidth]{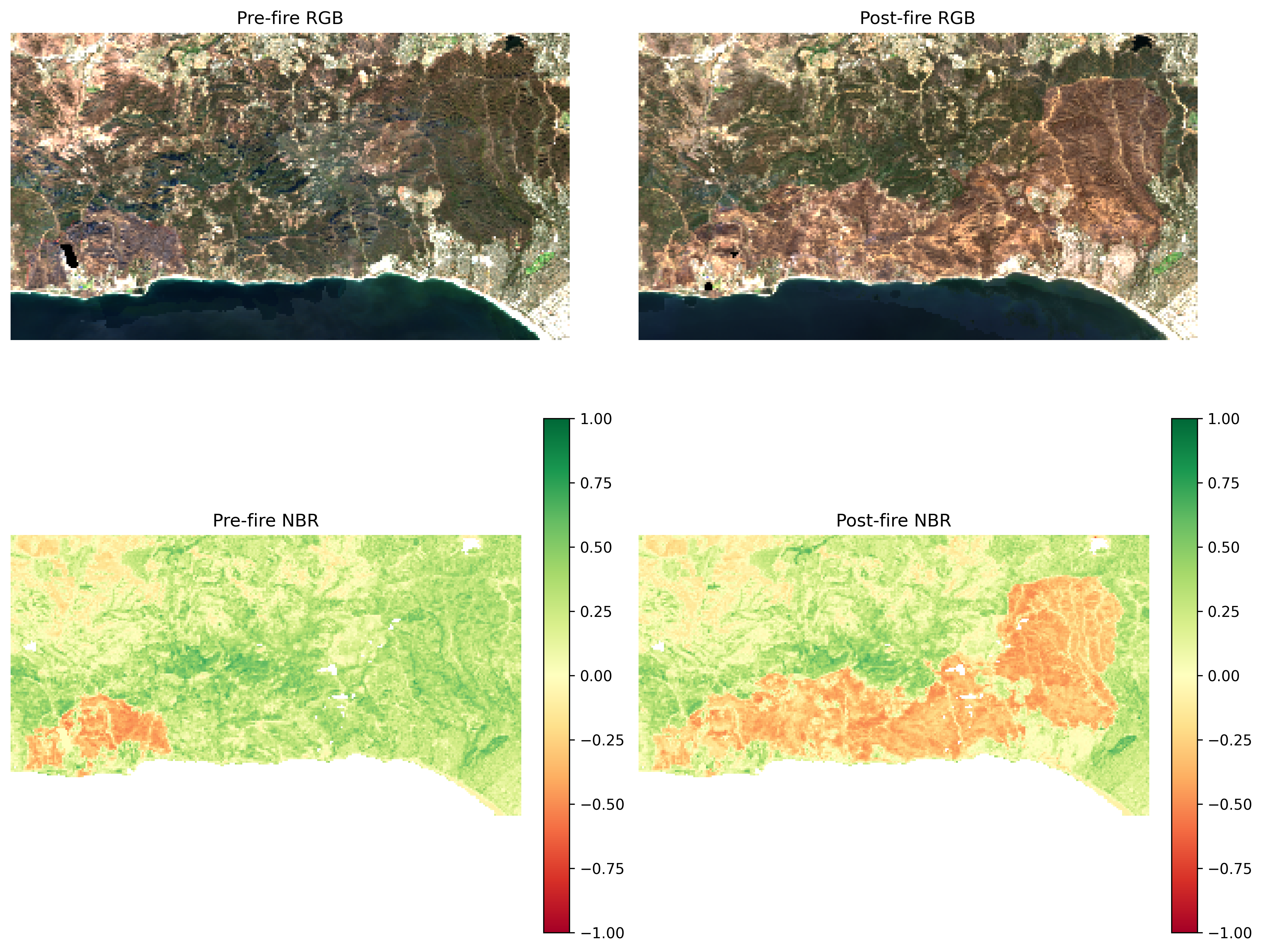}
        \caption{RGB and NBR imagery before and after the Pacific Palisades fire}
        \label{fig:rgb_nbr}
\end{figure}

\begin{figure}[ht]
    \centering
    \includegraphics[width=1.0\columnwidth]{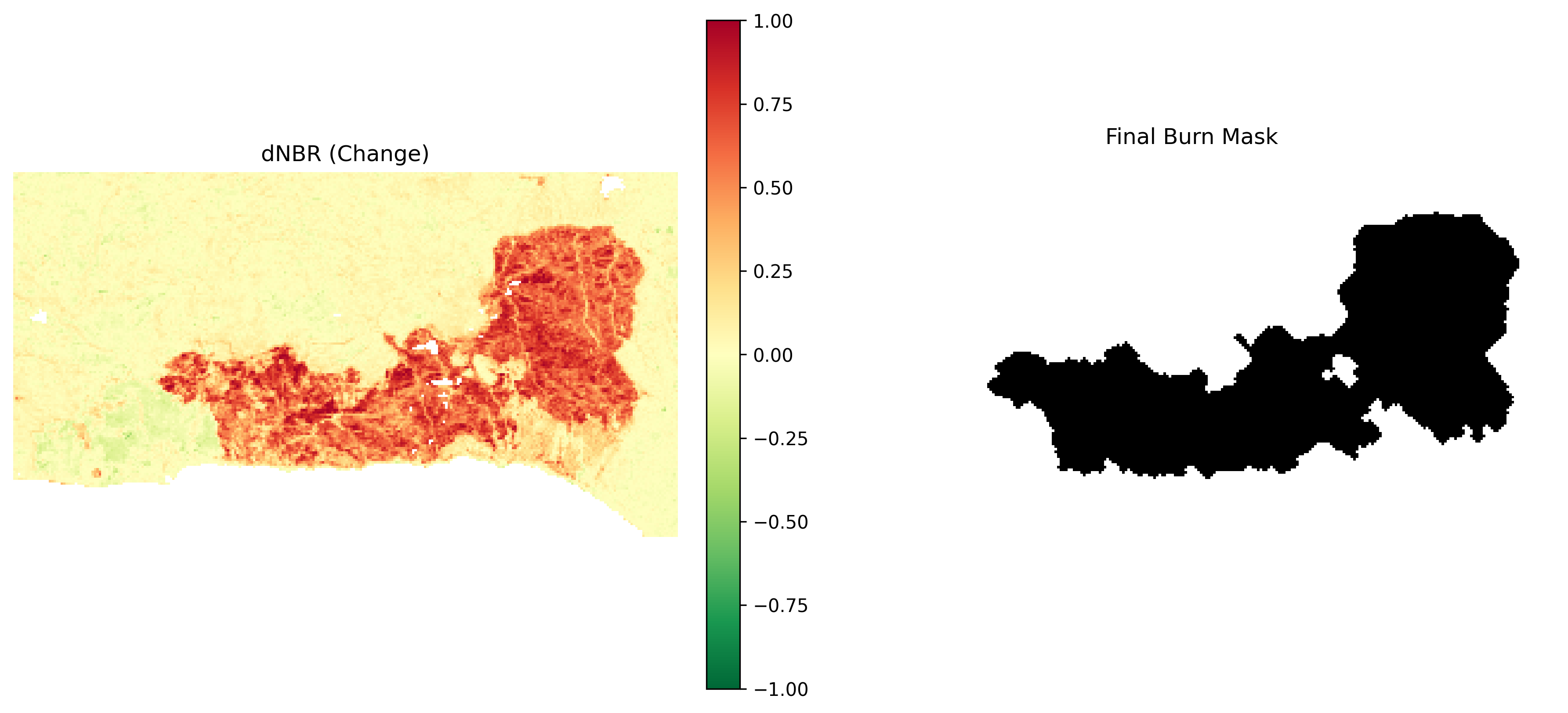}
    \caption{dNBR and burn mask results depicting the range and severity of the fire}
    \label{fig:dnbr_burnmask}
\end{figure}

Through analysis of the Sentinel-2 imagery we created a detailed map of the burn scar of the Pacific Palisades fire, first mapping the RGB and NBR as exhibited in \autoref{fig:rgb_nbr}. The two RGB photos at the top of the figure depict the study area before the fire and the immense scarring after the fire seen in the orange and black landscape; the two diagrams on the bottom present the same area in terms of NBR. The scale to the right of the NBR images shows the range of the NBR index, with negative values, shown as yellow, orange, and red, which are entirely burned, and positive values, shown as green, being healthy vegetation. The dNBR and a final burn mask were then mapped as seen in \autoref{fig:dnbr_burnmask}. The dNBR illustrates the impact of the fire, or the change in NBR by calculating the difference between the pre-fire NBR and the post-fire NBR. Calculated by dNBR with a threshold of 0.3, the raw burnt area was estimated to be 23,559 acres. The final burn mask was cleaned by applying morphological operations, creating a refined and contiguous burn scar. This cleaned mask served as the basis for consistent validation of the spread simulation through the confusion matrix.
\subsection{Parameter Optimization}
50 trials were run during parameter optimization which significantly improved the simulation's performance. After optimization, the best-performing set of parameters was identified. While future optimization could be more meticulous and exact, generalizing the model's capability offers a realistic examination of its performance ability. Each parameter assigns a random probability between 0 and 1 to the overall probability of the fire spreading to another cell. $p_{\text{medium}}$ and $p_{\text{dense}}$ determine the probability of spread through medium and dense vegetation respectively, while $c_{\text{wind}}$ and $c_{\text{slope}}$ assign a multiplicative factor for wind. The most optimal parameters of these trials were identified:
\begin{itemize}
    \item $p_{\text{medium}} = 0.312$
    \item $p_{\text{dense}} = 0.191$
    \item $c_{\text{wind}} = 1.246$
    \item $c_{\text{slope}} = 0.360$
\end{itemize}
The parameter values ranged greatly between trials, but these parameters were found to be the most ideal for general spread prediction.
\subsection{Confusion Matrix}
The confusion matrix that served as the foundation for all model validation is depicted in \autoref{tab:confusion_matrix}. Its success is clear in the large ratio of TP and TN as compared to FP and FN. In \autoref{tab:confusion_matrix} ``B'' refers to burned, and ``UB'' refers to unburned.
\begin{table}[ht]
    \centering
    \caption{Confusion matrix comparing simulated and observed cell classification}
    \label{tab:confusion_matrix}
    \begin{tabular}{ccc}
        \toprule
        & \textbf{Observed B} & \textbf{Observed UB} \\ % Shortened headers
        \midrule
        \textbf{Predicted B} & 7,203 (TP) & 4,703 (FP) \\ % Shortened headers
        \textbf{Predicted UB} & 1,173 (FN) & 21,405 (TN) \\
        \bottomrule
    \end{tabular}
\end{table}
\subsection{Burn Scar Validation and Model Performance}
\begin{figure*}[ht] 
    \centering
    \includegraphics[width=2.0\columnwidth]{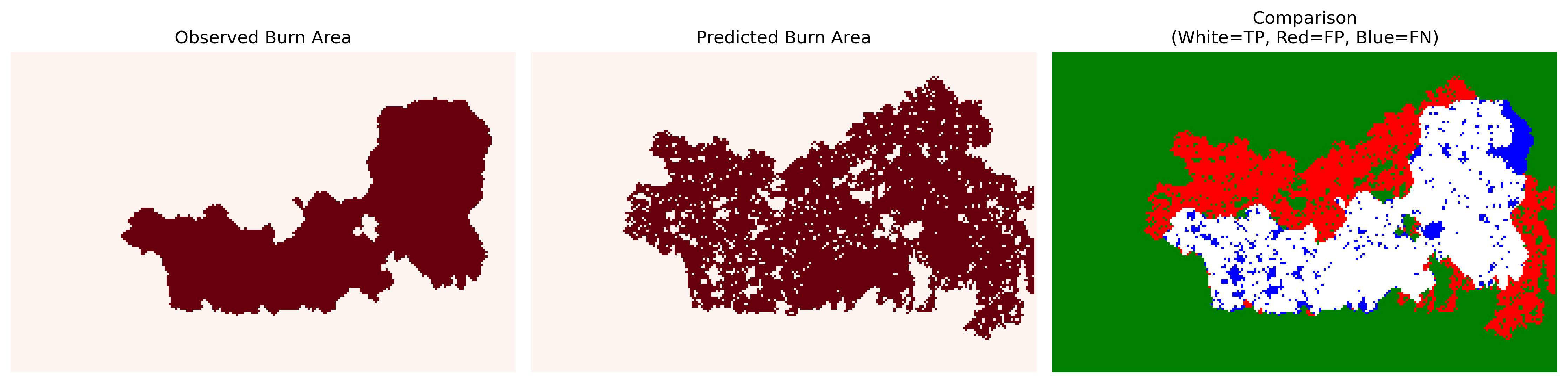} 
    \caption{Evaluation of the Palisades fire simulation}
    \label{fig:simulation_evaluation}
\end{figure*}

The simulated burn scar produced by the CA model we developed as shown in \autoref{fig:simulation_evaluation} was validated against the burn mask previously mapped. The observed burn area and predicted burn area illustrate strong agreement in core impacted areas, but show discrepancies in the eastern and southern regions where the model overestimated spread into unburned areas. To examine the simulation's success we calculated multiple validation metrics. The model's precision was 0.605, representing that 60.5\% of observed burned cells were agreeably predicted as burned. Its recall was higher at 0.860, indicating that 86.0\% of the observed burned cells were classified accurately by the model. Most importantly, the model's F1 score, a balance between precision and recall, was 0.711, showing the CA's real capability for fire prediction. These metrics suggest that while the model is successful in capturing observed burn areas, it often over-predicts burned regions, which is common in fire prediction models using similar variables. 
Additionally, the model's CA framework makes it highly operationally viable. The simulation, which focused on the primary period of 24 days during the fire, completed each simulation in an average of 1.22 seconds. This average was derived from the 50 parameter optimization trials, illustrating the model's optimal nature for real-time prediction and use in emergency-response contexts.
\section{Discussion}\label{sec:discussions}
The results of this study clearly show that a CA model relying on the variables of NDVI, wind, and slope, can predict fire spread to a realistic degree of accuracy while remaining simple enough to be applicable for real-time usage. Its F1 score of 0.711 demonstrates a strong match between simulated and observed burn areas, especially for such a simplified model with low computational demand. Its high recall of 0.860 indicates the model's exceptional ability to capture a large portion of the burned cells, while its lower precision of 0.605 shows its tendency to overestimate spread into unburned regions. This is very common in models that rely on simplified probability functions based on complex variables, as also seen by \cite{xu2022modeling} and \cite{zou2023attention}.

These findings align closely with previous research surrounding the applicability of CA-based modeling technology. While these approaches have low computational demands, their predictive accuracies come at a cost. Xu and colleagues illustrated that combining the structure and rules of CA with machine learning can produce more realistic results, it comes at the cost of higher data demands and less emergency response usability \citep{xu2022modeling}. Similarly, Zou et al. achieved greater realism with deep learning incorporation and remote imaging but faced the shared operational barrier of model training complexity and heavy reliance on high-resolution data \citep{zou2023attention}. In contrast to these two approaches, the presently developed model uses publicly available datasets and has limited reliance on high-resolution datasets and low computational demands, making it more ideal for informing real-time decisions concerning fire spread. This aligns with the theory of Crawl et al. that operational viability is arguably more important than raw accuracy in fire modeling \citep{crawl2017firemap}.

Beyond its evident statistical accuracy, the model's true strength and applicability lies in its rapid response time and public accessibility. Compared to advanced deep learning methods and large, complex datasets, such as those in previous studies, the CA framework developed in this study relies entirely on open-access data sources allowing the model to remain cost-effective and adaptable for broad use \citep{crawl2017firemap, karasante2025seasfire}. As emphasized by the work of Karasante et al., balancing operational viability and accuracy are crucial to applying fire modeling in operational settings. The simple CA framework allows for deployment in active fire areas where time and computational infrastructure are limited \citep{karasante2025seasfire}.

While effective in most core regions, the present model introduces limitations from its simplicity. The NDVI may have oversimplified the large diversity of fuels and moisture conditions, especially as seen in the urban, developed area of the Pacific Palisades. The model also doesn't account for changing weather or fuel conditions during a fire, and its wind variable is oversimplified, depending on an averaged hourly component. Additionally, the Palisades fire was addressed with a multitude of firefighting resources which is not incorporated into the current model. This resulted in an excess of false positives due to the firefighting protection of areas to the east and south of the Palisades. Addressing these limitations would require more dynamic environmental inputs at the cost of greater computational demands.

This work has broader implications. It demonstrates the value of operationally viable wildfire prediction models in public safety, offering a less data-heavy option to inform the public and firefighting efforts. Specifically, real-time prediction and visualization of fire spread could be used to inform more rapid and exact evacuation decisions, guide firefighting resources and physical efforts, and improve fire danger communication to the general public. Through the model's publicly accessible datasets, little computational demand, and rapid prediction time, it can be adapted for deployment not only by the government, but also by local agencies, NGOs, and community volunteer organizations.

Based on this study's findings, emergency management organizations would benefit from adapting to CA-based frameworks to inform their fire containment and aid efforts and complement data-heavy tools that have much greater computational and response time demands. This could include using simplified CA models to pre-load accessible datasets such as vegetation, wind speed and direction, and slope to inform rapid fire management preceding results from data-heavy models. Agencies could partner with remote imaging and satellite data providers to inform their own efforts alongside government action. In addition, the government could integrate these models into current prediction software to quicken public warning systems and their firefighting actions. Finally, the software framework could also be used publicly unlike many other models due to its publicly accessible datasets to provide affected communities with immediate realistic evacuation urgency, in turn improving evacuation compliance and reducing loss of life from wildfires.

In summary, an efficient, simplified CA model can fill in a critical gap between slow, data-heavy prediction models and overly simplistic inconsistent systems. The balance of the two offers operational viability and predictive accuracy and shows the potential to have a true impact on wildfire preparedness and burn alleviation.

\section{Conclusions}\label{sec:conclusions}

This study aimed to develop and evaluate a wildfire spread model balanced between operational simplicity and burn accuracy. The central goal was to determine if a CA model could deliver accurate prediction results while remaining efficient enough to be used for real-time emergency response efforts. Our results have shown that a model using the three key variables of NDVI, wind, and slope along with the standard spread rules of CA can achieve this original goal. In general, the results of the study and the evaluation metrics used demonstrate that the model not only functions, but yields a reasonable level of accuracy. Although its tendency to overpredict is a definite limitation, it is a common setback that comes from simplifying complex environmental processes to probabilities. Most importantly, the predicted burn scar aligned closely with the evaluated Palisades Fire in the core burn region, capturing its essential dynamics and key influences.

While its simplified design suits wildfire spread modeling, it comes with many limitations. The model's overprediction is the most notable limitation that likely stems from its inability to account for firefighting intervention and the large variety of fuel from urban construction in and around the Palisades. Additionally, the model's reliance on open-access datasets also results in many limitations. Low resolution data and highly simplified natural processes don't show the necessary dynamic nature of fire spread. Lastly, while the CA is effective for balancing low-level computing and high accuracy, it comes with limitations, such as simplifying large regions to relatively large cells.

Future research should aim to bridge this gap. Many studies have highlighted the importance of balancing accuracy and speed, and this theme must remain a focus \citep{crawl2017firemap, karasante2025seasfire}. While the framework of this study has proven effective and accurate, particularly in the context of informing emergency response, incorporating constant weather and firefighting inputs could improve the simulation's consistent overestimation. Most importantly, future research may pursue how models like this can be integrated into existing emergency response efforts. Wildfires are becoming more intense and common year after year, and our technology to confront them must also adapt. This research shows that simplified wildfire spread models can provide valuable information supporting fire mitigation efforts despite their lack of heavy data and signifies a step toward more accessible and practical tools for fire management.

\bibliographystyle{elsarticle-num}

\bibliography{citations} 

\vspace{1cm}

\end{document}